\documentclass[conference]{IEEEtran}

\IEEEoverridecommandlockouts

\usepackage{xcolor}
\usepackage[pdftex]{graphicx}
\graphicspath{{./Figures/}}
\DeclareGraphicsExtensions{.pdf}
\usepackage{wrapfig}
\usepackage{url}

\begin{document}
\title{IoT Vulnerability Data Crawling and Analysis%
\thanks{%
\protect\begin{wrapfigure}[3]{l}{.9cm}%
\protect\raisebox{-12.5pt}[0pt][0pt]{\protect\includegraphics[height=.8cm]{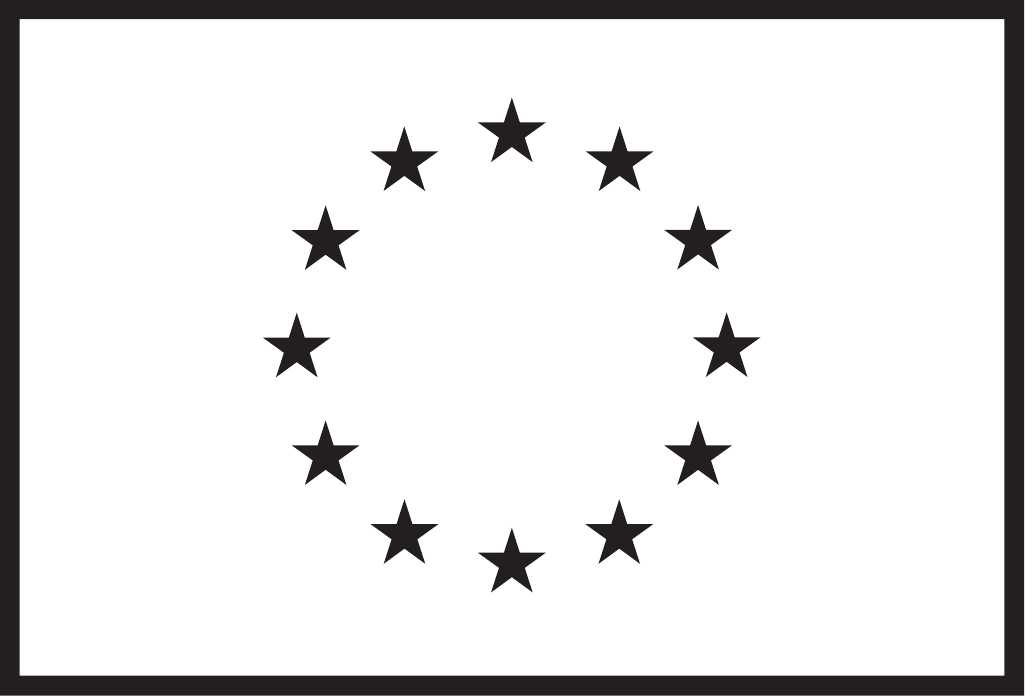}}%
\protect\end{wrapfigure}%
This project has received funding from the European Union's Horizon 2020 research and innovation programme under grant agreement no. 786698. The work reflects only the authors' view and the Agency is not responsible for any use that may be made of the information it contains.}}

\author{\IEEEauthorblockN{%
Stavros Shiaeles\IEEEauthorrefmark{2},
Nicholas Kolokotronis\IEEEauthorrefmark{1},
Emanuele Bellini\IEEEauthorrefmark{3},
\vspace*{4pt}}

\IEEEauthorblockA{\IEEEauthorrefmark{2}University of Plymouth, UK. Email: sshiaeles@ieee.org}
\IEEEauthorblockA{\IEEEauthorrefmark{1}University of Peloponnese, Greece. Email: nkolok@uop.gr}
\IEEEauthorblockA{\IEEEauthorrefmark{3}Mathema s.r.l., Italy; Khalifa University, UAE. Email: emanuele.bellini@ieee.org}

}
\maketitle


\begin{abstract}
Internet of Things (IoT) is a whole new ecosystem comprised of heterogeneous connected devices ---i.e. computers, laptops, smart-phones and tablets as well as embedded devices and sensors--- that communicate to deliver capabilities making our living, cities, transport, energy, and many other areas more intelligent. The main concerns raised from the IoT ecosystem are the devices poor support for patching/updating and the poor on-board computational power. A number of issues stem from this: inherent vulnerabilities and the inability to detect and defend against external attacks. Also, due to the nature of their operation, the devices tend to be rather open to communication, which makes attacks easy to spread once reaching a network. The aim of this research is to investigate if it is possible to extract useful results regarding attacks' trends and be able to predict them, before it is too late, by crawling Deep/Dark and Surface web. The results of this work show that is possible to find the trend and be able to act proactively in order to protect the IoT ecosystem. 

\end{abstract}


\section{Introduction}

The grand vision of the Internet of Things (IoT) \cite{c4} is to establish a whole new ecosystem comprised of heterogeneous connected devices ---computers, laptops, smart phones, and tablets, as well as, embedded devices and sensors--- that communicate to deliver environments making our living, cities, transport, energy, and many other areas more intelligent. The current forecasts on the number of connected IoT devices is expected to exceed the number of mobile phones \cite{c1}.

As the networked devices become ubiquitous, cyber-attacks become more frequent and even more sophisticated. In fact, IoT devices may contain confidential and private information, and many security threats have emerged that aim to exploit the weaknesses of current IoT infrastructures \cite{c3}. There are already numerous recent examples of cyber-attacks that exploit the Internet-connected appliances, such as smart TV, cameras, vehicles, in order to perform denial-of-service (DoS), distributed DoS (DDoS) attacks, spy on people during their daily life, and hijack communication links taking full control of devices that are remotely controlled such as drones, robots and vehicles. Computer-controlled devices in automobiles, like locks, brakes, and engines, have been shown to be vulnerable to numerous attacks. These devices are currently not connected to the Internet, and hence are less vulnerable to remote attacks. If however are accessed remotely then attacks' impact can be severe leading even to life-threaten situations such as steering wheel while driving, insulin pumps malfunctions, CT-scanners image modification, implantable defibrillators malfunction, and x-ray systems manipulations. Security issues are rising also in Critical Infrastructures (CIs) such as oil streams, transport, energy, water that are deploying IoT for monitoring and operation automation. Cyber-security attacks targeting CIs are expected to have a significant negative impact at economical and societal level in the next decade \cite{c2}. In fact, according to the World Economic Forum (WEF), cyber-attacks need to be considered a global risk, and are ranked at the 5th place for likelihood and 7th place for impact \cite{a25} in the WEF top 10 Global Risks. In this respect, the Cyber-Trust project aims at providing a platform equipped with novel tools and methods so as to allow the prevention, detection, and mitigation of large-scale cyber-attacks in IoT. This work is interwoven with the project aiming to a proactive working system based on the publicly available information in Deep/Dark and Surface web.

The article is organized as follows: in section \ref{relatedwork} the state of the art related to vulnerability assessment is provided; in section \ref{crawler} the implementation characteristics for the  vulnerability crawler are presented; in section \ref{dataanalysis} a preliminary analysis of the data about vulnerabilities is provided; section \ref{discussions} is devoted to discuss the impact of the proposed solution in the IoT domain and the potential of an even more sophisticated analysis; conclusions and future work are reported in section \ref{conclusions}.


\section{Related work}
\label{relatedwork}

Since vulnerabilities are becoming more prominent, many researchers and companies focused their efforts on strengthening the global protection against vulnerabilities.

In 2015, a team of researchers built a Cross-site Scripting (XSS) vulnerability detector \cite{a4}. The XSS vulnerability detector runs a crawler that can interpret Javascript code, and can detect if a website is protected against XSS vulnerability. The crawler provides a percentage of potential targets for Cross-Site Scripting and helps these targets to upgrade their protection.

Since Injections are one of the most powerful attacks according to OWASP top 10 security risks \cite{a1}, researchers in \cite{a5}, \cite{a6} tried to simulate realistic injection attacks scenarios to measure the protection of websites. Others works focused on the response of Injections to measure the resistance against them \cite{a7}, \cite{a8}. An interesting approach comes with a tool called ``Sania'' \cite{a9}. In this work, researchers directly intercept the query between the website and the database, and detect if the query is an attack or not, allowing it to detect the weak spots of the database. After experiments on real-world web applications, they found that their solution was more efficient than a regular scanner.

A black-box web application vulnerability is presented in  \cite{a10}, \cite{a11}. The work is able to test and analyze the resistance of a website against vulnerabilities, including, but not limited to, Cross-Site Request Forgery, Cross-Site Scripting and Injection. 

A similar approach to find vulnerabilities with the use of a crawler has been introduced in \cite{a14} and \cite{a15} using Dark Web. However, these works were for drug sales and illegal activities respectively but show that crawlers could be an important component in retrieving critical information from forums \cite{a12} or on any side of the Dark Web \cite{a13}.

\section {Vulnerability}
Vulnerability is defined according to ISO 27005 \cite{a16} as ``A weakness of an asset or group of assets that can be exploited by one or more threats'' and according to IETF RFC 4949 \cite{a17} as ``A flaw or weakness in a system's design, implementation, or operation and management that could be exploited to violate the system's security policy''. ENISA \cite{a18} provides a more comprehensive definition which is ``The existence of a weakness, design, or implementation error that can lead to an unexpected, undesirable event compromising the security of the computer system, network, application, or protocol involved''.  

Knowing the characteristics and behavior of devices in an IoT infrastructure is a necessary requirement for an effective implementation of cyber security and resilience strategy \cite{a20}, \cite{a23}. In this work we identified the most critical class of vulnerabilities for IoT infrastructure whose information is gathered by the crawler. These categories has been defined as follow:

\begin{itemize}
\item \textit{Arbitrary Code Execution}: The term ``Arbitrary Code Execution'' is used to describe the execution of any kind of code on a target that could be either physical like a machine, or virtual like a website. In order to differentiate every vulnerability, in the rest of this paper, the term ``Arbitrary Code Execution'' will denote any vulnerability that is not already defined: we consider this vulnerability in the ``Others'' section; e.g., an Injection is an Arbitrary Code Execution, but on this report, we will consider them separate.

\item \textit{Injection}: An Injection is an attack allowing hackers to execute malicious code into a system, one of the most common being the SQL Injection. This vulnerability generally allows the hacker to see dynamically generated parts of the website not intended to be displayed that way, or simply access the database of the website. This vulnerability is probably one of the most dangerous, since there are many different kinds of Injections, and protecting a website against all of them could be quite challenging, and letting a hacker run an Injection could destroy companies, especially those keeping sensitive data.

\item \textit {Broken Authentication}: This vulnerability concerns every website with a login system. A Broken Authentication simply allows a hacker to login to a website without having an account. Once logged in, the hacker can potentially change the password and make it inaccessible to the owner, and retrieve the data stored in the account. The method used can vary, but a brute force attack is the easiest way: it consists of trying every password possible on a really high frequency.

\item \textit {Cross-Site Request Forgery}: Also known as CSRF, this attack targets directly a person; the hacker tricks the victim into sending a malicious request to a regular website. That way, the website has no way of knowing that the request has been designed by a hacker. The request sent has to change a state, such as changing a password or purchasing something, because even if the request retrieves data, the hacker cannot see the response, since only the victim is logged in.

\item \textit {Server-Side Request Forgery}: Also known as SSRF, this attack targets a website. On certain websites, it is possible to retrieve information by changing URLs, that means that the hacker can potentially write a full request in the URL, changing completely the intended use. The server will try to read the URL given, and if it is well designed by the hacker, the server will give back hidden information.

\item \textit {Cross-Site Scripting}: Also known as XSS, the Cross-Site Scripting is a type of vulnerability that allows the hacker to run scripts on trusted websites. Therefore, when a user tries to visit the website, he has no way of knowing that it has been infected, and will run the script. That way, the hacker can have access to the user’s data stored in the browser, such as passwords or credit card information.

\item \textit {Remote Code Execution}: A Remote Code Execution is simply an Arbitrary Code Execution from one machine to another, via Internet most of the time.

\item \textit {Remote Command Execution}: Similar to the Remote Code Execution, this vulnerability executes a malicious command from one machine to another.

\item \textit {Denial of Service}: This attack has the goal of making a website or a server unavailable, for example by sending a very large number of requests. This vulnerability does not have a purpose of stealing information, but could be used to weaken a website while executing malicious code.

\item \textit {Buffer Overflow}: This attack, generally known by programmers since it actually is a common error, has the purpose of sending meticulously written code as an input to a website, for example a basic e-mail field. A not prepared website could potentially lose its database, and even lose the control of everything.

\item \textit {Privilege Escalation}: A Privilege Escalation is simply the upgrade of the privileges of one user on a website. This weakness could be the repercussion of another attack, and could be potentially really dangerous for a website if the hacker can upgrade himself into an administrator.

\item \textit {Arbitrary File Manipulation}: An Arbitrary File Manipulation is generally an upload of malicious files into a website, allowing the hacker to make a first step in the attack of the website. These files generally contain malicious code, and the hacker only has to find the way to execute it once uploaded on the website.

\item \textit {Directory Traversal}: This vulnerability, also known as Path Traversal, has the purpose of accessing files hidden in the website that are not accessible. Depending of the information the hacker could find in them, this vulnerability could potentially be really dangerous for the website.

\end{itemize}

\subsection {Dark web as a data source}
There is a plethora of vulnerabilities that affect devices in an IoT infrastructure and is important to have timely information about them in order to activate appropriate risk-reduction countermeasures (e.g. software upgrade). To achieve that, we need to gather information about those vulnerabilities from many sources, and especially from markets where cyber-criminal reside. Due to the fact that most of the markets are dealing with illegal activities, are located in place that require specific software that use encryption and anonymity in order to be accessed and this web is called Dark Web. The web is presented as three layers: the top layer is the regular web, where can be access with a simple browser and well-know websites such as Facebook, Twitter, etc exists. Then comes the Deep Web, which is only the part of the web that cannot be accessed by using regular search engines. For example, when you are checking your mails, this is a page you cannot directly access by searching of Google; because it is part of the Deep Web. 

Finally, the last part, often mixed with Deep Web: the Dark Web. The Dark Web is actually a part of the Deep Web that utilised anonymizing software, most common one being "Tor", where enter a link ending with the extension ".onion" which are links to Dark Web pages.  On those pages, you can find most of the illegal activities on the Internet such as weapon sales, drug sales as well as vulnerabilities. 

Finding information on the Dark Web is not an easy task, so we will need a tool to inspect multiple markets at a time and this tool would be a web crawler. A crawler, also known as spider, is a program which inspects contents of a given website and finds pre-specified data. However, the program has to know the website it is crawling and specific regular expression should be implemented based on the structure of the website. Thus the crawler should be modular and each website should have its own module in the developed crawler system for scalability.

Python language is famous for having a lot of libraries available for web crawling, so our final goal would be to build a web based system that will run Python crawlers on different Dark Web markets. A good resource for selling vulnerabilities spotted is called 0daytoday\cite{a2} and it was utilised in our developed system as well as other websites to retrieve vulnerabilities. These vulnerabilities will further analysed to the results would be displayed on the web frontend of the developed web based system so that we could have a better understanding of the kind of vulnerabilities the hackers are using and predict the trend. To be more specific, our interested resides mostly onvvulnerabilities called "0day" which are vulnerabilities that has just been found and the vendor has not provided any patch. This means that the malicious code has more chances to succeed infect a system.

Furthermore, even though it is possible to find powerful vulnerabilities for famous websites or software on the Dark Web, we can also find minor vulnerabilities on regular websites, which could either be specialized on 0day vulnerabilities \cite{a2}, or just be websites on social network or used from many groups such as Twitter and Pastebin. This means that our website will have to run scripts for Dark Web markets, but also for regular websites in order to collect the maximum amount of data possible to do the correlation and export meaningful results.

\section{Crawler Implementation }
\label{crawler}
The implementation of a Crawler for digital resources on the web requires several precautions. The main criteria to be considered are: a) the time needed for the harvesting that depends to the computational power,  bandwidth available, answering time of the server contacted; and b) type of the resources collected in term of volume, complexity (e.g. web page with a complex structure to be processed). Because of the vulnerability and related information are published on web pages of the Dark Web in a not standard/structured format differently to what happens in Open Archive domain where the harvesting is supported by the OAI-PMH protocol\cite{a21}, the Crawler has to be implemented in a clever manner in order to avoid the management of each single case.  

First, we need to build a functional crawler: for this we will use the library "selenium" which allows us to simulate the navigation on a website by using a "headless browser". A headless browser is a regular browser, where you can normally access pages, you just can’t see the window open: this makes the program run much faster. With selenium, we can easily get the HTML code of any website, which is essential to find data. We will also need a library to easily read HTML code: a famous Python library called BeautifulSoup will help us with that; it allows us to navigate through the code like it was a tree. With the knowledge of how the website is built, we can search for specific tags and find the exact piece of information we need.

After successfully data collection from the website, data needs to be processed. Thus, search for specific keywords is important in order to classify the data in the vulnerability categories that has been defined in Section 2. For simple vulnerabilities like Injections, only simple keywords are utilised such as "Injection" or "SQL", but for XSS more things should be considered like the use of capital letters or only lowercase or the use of the full name “Cross-Site Scripting”.

The next step after the classification of vulnerability found storing the results into a database is important for keeping history and been able to analyse them and export meaningful results. A MySQL database was used for storing data and Python for the SQL requests to the database. The fields defined in database for every new vulnerability that found, are the type, the user who found the vulnerability, the date, hour, and the website where the vulnerability has been found. 

Finally, we need to create a website that runs our Python scripts, gather every vulnerability we find on every website, and present them in the form of charts. Since we used Python scripts, we will use Django to create our website. Django is an open source web application framework written in Python, and it allows us to run Python scripts easily in our website. In this website, there would also be a 2-factor authentication system in order to avoid brute force attacks, we will use Google reCAPTCHA for that.

The script launches would be divided into two parts: first, the script should load the entire page, then see if any vulnerability has never been stored in our database. In order to do that, we just need to compare the date of the last vulnerability stored in the database, then take everything with a later date. Then, the script will just wait and refresh the page every minute to see if any new vulnerability has been detected, so we will just keep comparing dates with the last one stored in our database. 

We would need to run three different scripts: one on 0daytoday to gather Dark Web vulnerabilities, one on Exploit-DB to confirm the vulnerabilities we found on the Dark Web, and maybe find different   kinds, and finally on Twitter to have a different approach, for example if someone finds a vulnerability but does not want to store it in a database, this script would still find it. There should also be a way to see the scripts up-time on the website. Also, the scripts should launch automatically: since these scripts require a functional Internet connection, they can stop if the Internet connection is no longer available. The user should not need to launch the scripts again every time the connection is not strong enough. Plus, since the scripts will most likely find the same vulnerabilities, there should be a way of checking the database and see if a vulnerability of the same type had already been stored. For that, every time a vulnerability would have been added, the database retrieves the list of the vulnerability of the same type from two months ago, and compare word by word the matching parts. If there are too many words in common with another vulnerability (which is likely to happen since we have three different sources dealing with the same vulnerabilities), the vulnerability is not added.

To display the results, we will use an open source library called Chart.js\cite{a19}. We would need to make one chart for every month, and display the vulnerabilities in a way that we can see the global evolution of vulnerabilities, but we would also need to see the evolution of each of them. The clearest way to do so would be with bars: we could see the total number of vulnerabilities detected per day with the height of a bar, and we could separate each vulnerability with different colors. If we use different colors, we would need to add a legend, and this legend could add an option of hiding some vulnerabilities and only displaying the ones the user wants.

Finally, there should be a way to see what are the last vulnerabilities that have been analyzed by the program, so for each type of vulnerability, a list of the 50 last vulnerabilities found would be displayed on the website for demonstration purposes. These steps are summarized on the Fig. 1.

\begin{figure}[t]
\centering
\includegraphics[width=\linewidth]{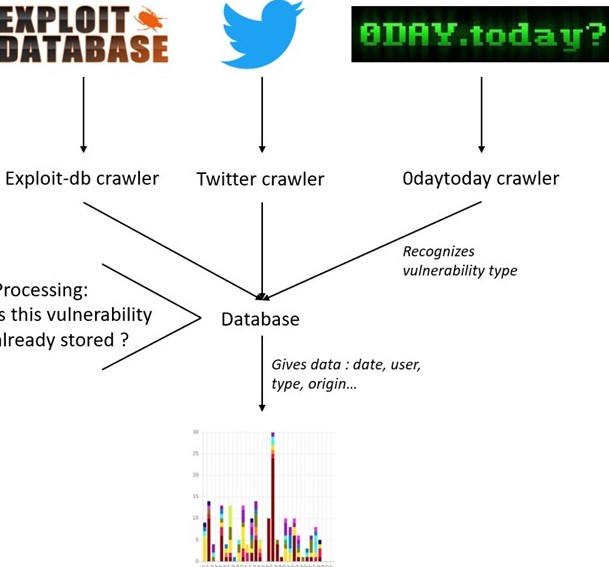}
\vspace{-1em}
\caption{Proposed Method}
\label{fig.platform}
\vspace{-0.5em}
\end{figure}

\section{Data Analysis}
\label{dataanalysis}

The first goal was to run the three scripts from the month of January 2018 until now (the end of month of August 2018) to have enough data to process. From these data, we would be able to build our charts, and try to understand what the most common vulnerabilities are. Since the vulnerabilities are displayed with their percentage on the website, it is easy to check the most common ones for each month. 

Using OWASP top 10\cite{a1}, initially we wanted to compare our results with their top 10 list in order to see even in this short period of time if there are any commonalities. Based on the list the first major vulnerability is the Injection and Table 1 shows our results for the  the three most common vulnerabilities percentages per month.

\begin{table}[]
\centering
\begin{tabular}{|p{1.1cm}|p{1.9cm}|p{2cm}|p{2.2cm}|}
\hline
Month (2018) & 1st                & 2nd                  & 3rd  \\ \hline
 January     & Injection (28\%)     & XSS (11\%)            & Rem. Code Exec.(11\%)    \\
 February    & Injection (36\%)     & XSS(16\%)             & Remote Code Exec.(9\%)  \\
 March       & Privil. Escal.     & Injection (13\%)        & XSS (13\%)      \\
 April       & XSS (24\%)           & Rem. Code Exec.(15\%) & Buffer Overflow (13\%)        \\
 May         & Injection (24\%)     & XSS(22\%)             & CSRF (16\%)\\
 Jun         & Injection (24\%)     & CSRF(15\%)            & Buffer Overflow (15\%)  \\
 July        & DoS (21\%)           & Injection (14\%)      & XSS (13\%)  \\
  August     & DoS (30\%)           & XSS(14\%)             & Buffer Overflow (11\%)\\\hline
\end{tabular}

\caption{top 3 most common vulnerabilities per month in the given time window }
\vspace{-3em}
\end{table}

From the tables can be concluded that our list and OWASP agreed in terms of  the most common vulnerability which is Injection as it can be observed in every month in our top three, and four times first in our list. It is also worth mentioned the existence of Cross-Site Scripting vulnerabilities in our top list at each month. 
Another useful information is the presence of Denial of Service vulnerability on the top of the list two months in a row, even though it was not even on the top 3 before that. 
To analyze the implication of all other vulnerabilities, the next chart will show the part of each vulnerability per month.

\begin{figure}[t]
\centering
\includegraphics[width=0.8\linewidth]{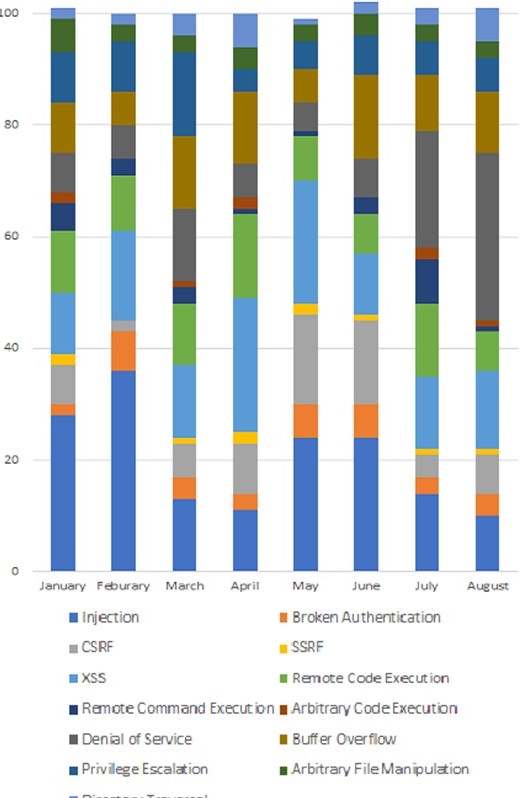}
\caption{Evolution of each vulnerability per month in the given time window}
\label{fig.platform}
\vspace{-1.5em}
\end{figure}

We can see that most of the time, Injections, Cross-Site Request Forgery, Cross-Site Scripting, Denial of Service and Buffer Overflow are the most common vulnerabilities.

Now that we have the possibility to see what the most common vulnerability was for each month, we can also wonder why a vulnerability would be on the top of the list.
Let us take the example of the month of February, where the percentage of Injection was 36\%, against only 16\% for the second of the list, which was Cross-Site Scripting. The chart 2 presented later is the evolution of each vulnerability during the month of February.
We can see that around the middle of the month, there is a huge peak of Injections, almost 20 Injections on the same day, and on the day before that, we also find 10 Injections on that day. There could be many reasons for these peaks, but one of the most likely to happen is the massive breach in an application that leads to a lot of different types of Injection.

\begin{figure}[t]
\centering
\includegraphics[width=0.8\linewidth]{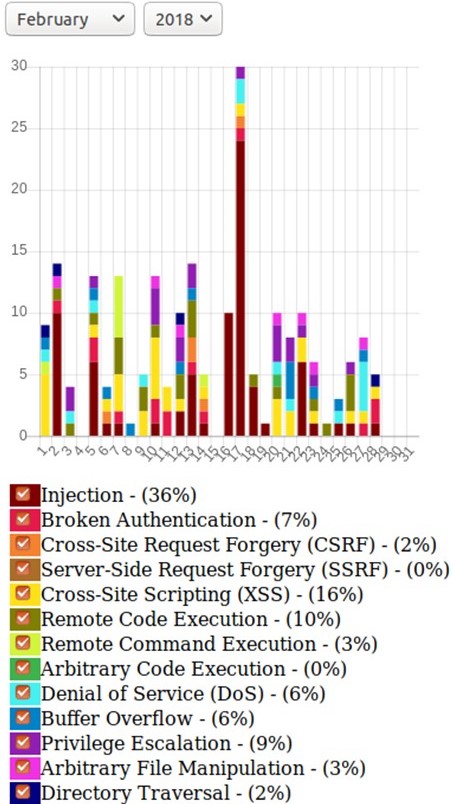}
\caption{Evolution of each vulnerability during the month of February}
\label{fig.platform}
\end{figure}

\section{Discussion}
\label{discussions}
The main goal of this work was to detect the most common vulnerabilities in order to have a better protection against them. Each month, we saw the evolution of each vulnerability in comparison with the others, and this led us to conclude that the most dangerous were the most common. 
However, we need to keep in mind that some of these vulnerabilities might be more dangerous than their ranking made us believe. For example, the Cross-Site Request Forgery used to be one the most dangerous vulnerabilities, but why such an important vulnerability would only appear one time on the table presented earlier? The answer is: the protection against this vulnerability is stronger than before. Right now, most of the frameworks have an anti-Cross-Site Request Forgery included. That means that even without thinking about protecting ourselves against this kind of vulnerability, we would be protected. However, without the use of a framework, an anti-Cross-Site Request Forgery is important, and we could even say essential.

Finally, we need to keep in mind that our website could also be a potential threat for hackers, since the data contained in our database could be quite long to gather. We are most likely protected against Broken Authentication since we used a Captcha for our login system, and the framework we are using, Django, has a lot of protection against the most common vulnerabilities included (anti-Injection, anti-Cross-Site Request Forgery, anti-Cross-Site Scripting…), but we need to remember that vulnerabilities could potentially target anyone.
Thinking that we are safe is not enough to be protected, and even it generally "happens to the others", once a website is hacked, it is really difficult to take the control back. We just need to see that even companies like Windows or Apple are really often targets of vulnerabilities to understand that it is crucial to defend ourselves against the most common attacks.

\section{Conclusions}
\label{conclusions}
The aim of this research was to investigate various resources of vulnerabilities in Deep/Dark and Surface web and develop a crawler that will be extensible and collect this information. This was achieved by developing a modular crawler that is accepting  websites in Deep/Dark and Surface web as modules. Furthermore we utilised the information collected and investigate the vulnerability trends by cross-validating them with well-known reports such as OWASP top 10. Results show that even in short time it is possible to get vulnerabilities' trends. Future extension of our proposed model would be the addition of events related with the vulnerabilities in order to calculate the impact as well as machine learning that would utilise the information and being able to predict the risk. This can be further extended and raise the level of logging for vulnerable IoT devices in order to act proactively and reactively in the case of attack, providing the necessary information to forensics examiners. Moreover, the vulnerability assessment enabled by the results of the present work, can be used as a basis to calculate the probability of infection of a device within a network in the emerging epidemic spreading approach of cyber attacks analysis \cite{a22}, \cite{a24}. In fact, knowing the nature of the vulnerability existing in a IoT infrastructure,  reduces the level of  uncertainty of the analysis result derived by a pure stimulative investigation.

\end{document}